\documentclass[12pt,a4paper]{article}

\usepackage[cp1251]{inputenc}
\usepackage[T2A]{fontenc}
\usepackage[english]{babel}

\pdfoutput=1

\newcommand{\MSbar}{\overline{\rm MS}}
\newcommand{\DRbar}{\overline{\rm DR}}

\begin{document}
\title{Riemann $\zeta(4)$ function contributions to $O\left({\alpha_s}^5\right)$ terms of Adler D-function
and Bjorken polarized sum rule in $SU\left(N_c\right)$ QCD: results and consequences}

\author{
I.~O.~Goriachuk\footnote{io.gorjachuk@physics.msu.ru}\\
{\small{\em Moscow State University,}}\\
{\small{\em Faculty of Physics, Department of Theoretical Physics,}}\\
{\small{\em 119991, Moscow, Russia}},\\
\\
A.~L.~Kataev\footnote{kataev@ms2.inr.ac.ru},\\
{\small{\em Institute for Nuclear Research of the Russian Academy of Science,}}\\
{\small{\em 117312, Moscow, Russia}}.\\
}

\maketitle

\vspace*{-14.0cm}
\begin{flushright}
INR-TH-2020-034
\end{flushright}
\vspace*{12.0cm}

\begin{abstract}
Two renormalization group invariant quantities in quantum chromodinamics (QCD), defined in Euclidean space,
namely, Ad\-ler D-function of electron-positron annihilation to hadrons and Bjorken polarized deep-inelastic scattering
sum rule, are considered.
It is shown, that the 5-th order corrections to them in $\MSbar$-like renormalization prescriptions,
proportional to Riemann $\zeta$-function $\zeta\left(4\right)$, can be restored by the transition to the C-scheme,
with the $\beta$-function, analogous to Novikov, Shifman, Vainshtein and Zakharov exact $\beta$-function in
$\mathcal{N}=1$ supersymmetric gauge theories.
The general analytical expression for these corrections in $SU\left(N_c\right)$ QCD is deduced
and their scale invariance is shown.
The $\beta$-expansion procedure for these contributions is performed and mutual cancellation of them
in the 5-th order of the generalized Crewther identity are discussed.
\end{abstract}

\section{Introduction}
The existence of the all-loop expression for the renormalization-group $\beta$-function is one of attractive properties
of $\mathcal{N}=1$ supersymmetric gauge theories \cite{Novikov:1983uc}.
This expression was called <<Novikov, Shifman, Vainshtein and Zakharov (NSVZ) exact $\beta$-function>>.
In the case of $\mathcal{N}=1$ supersymmetric Yang-Mills theory without additional matter superfields it has
geometric-progression related form, unambiguously defined by its first two coefficients $\beta_0$ and $\beta_1$.
The renormalization prescription, for which this exact formula is valid is theoretically distinguished.
In this NSVZ scheme the renormalized operators, corresponding to the axial and conformal anomalies,
lie in the same supermultiplet \cite{Jones:1983ip}.

Only thirty years later the renormalization procedure, providing the exact NSVZ relation
in all orders of the perturbation theory (PT), was formulated
for $\mathcal{N}=1$ supersymmetric quantum electrodynamics (SQED) \cite{Kataev:2013eta}.
This procedure implies that the higher derivatives (HD) regularization \cite{Slavnov:1972sq,Slavnov:1971aw}
in its supersymmetric form \cite{Krivoshchekov:1978xg} is employed and
the subtraction scheme is based on imposing special boundary conditions on the renormalization constants,
realizing the principal of minimal subtraction of logarithms (MSL).
It has been shown recently, that the use of the on mass shell (OS) subtraction scheme in $\mathcal{N}=1$ SQED
also provides the validity of the NSVZ formula \cite{Kataev:2019olb}.

First multiloop calculations of quantum corrections in supersymmetric models were conducted in the $\DRbar$ scheme
(originally applied in quantum chromodynamics in \cite{Altarelli:1980fi}, where it was called $\MSbar$ (reduction)),
employing the dimensional reduction for the regularization \cite{Siegel:1979wq,Capper:1979ns,Avdeev:1981vf}.
This scheme is analogous to the $\MSbar$ procedure \cite{Bardeen:1978yd}, which is used in non-supersymmetric theories
when the dimensional regularization \cite{tHooft:1972tcz} is introduced.
In the $\DRbar$ scheme the NSVZ relation is not valid. This was first understood as the result of analytical 3-loop
calculation.
However, as was shown in \cite{Jack:1996vg}, it is possible to restore the NSVZ relation by a special coupling constant
finite renormalization.
This fine tuned $\DRbar$ + (special finite renormalization) result can be reformulated by applying MSL-like
boundary conditions in the 3-loop approximation at least \cite{Aleshin:2016rrr}.

It was shown in \cite{Goriachuk:2018cac} that there is a special class of renormalization schemes in $\mathcal{N}=1$ SQED
for which the exact NSVZ $\beta$-function is valid in terms of the renormalized coupling constant.
It includes the three above mentioned renormalization procedures: HD + MSL \cite{Kataev:2013eta},
HD + OS \cite{Kataev:2019olb}, DRED + finite renormalization \cite{Aleshin:2016rrr}.
Particular schemes of this class are related with each other by finite renormalizations,
which form a subgroup of general renormalization group transformations \cite{Goriachuk:2019aqy}.
In the case of $\mathcal{N}=1$ SQED, this subgroup is non-commutative \cite{Goriachuk:2020wyn}.

In non-Abelian $\mathcal{N}=1$ supersymmetric gauge models the HD + MSL renormalization prescription
also ensures the validity of the NSVZ relation exactly in all orders of the PT.
It was shown in \cite{Stepanyantz:2020uke} and confirmed by numerous explicit calculations of supergraphs
(see, e.g., \cite{Kazantsev:2020kfl}).
In $\mathcal{N}=1$ supersymmetric Yang-Mills (SYM) theory without matter there is also a class
of renormalization schemes, in which the NSVZ formula is valid \cite{Goriachuk:2020wyn}.
The members of this class differ from each other by the choice of the normalization scale $\mu$.
They are related by the transformations of the commutative subgroup of general renormalization group transformations
\cite{Goriachuk:2020wyn}.

The exact expression for the $\beta$-function of quantum chromodynamics (QCD), similar to the NSVZ $\beta$-function
in $\mathcal{N}=1$ supersymmetric Yang-Mills theory without matter superfields in its form, is an attractive feature
of a special C-scheme.
In this renormalization scheme, proposed in \cite{Boito:2016pwf} for the investigation of scale-scheme ambiguities
in theoretical expressions for physical quantities in QCD, the corresponding $\beta$-function has the following form:
\begin{equation}
\beta^C\left(a_s^C\right) = \mu_C^2 \frac{\partial a_s^C\left(\mu_C^2\right)}{\partial\mu_C^2} =
\frac{-\beta_0 \left(a_s^C\right)^2}{1 - \left(\beta_1/\beta_0\right) a_s^C}. \label{eq:betaC}
\end{equation}
Only the values of the coefficients $\beta_0$ and $\beta_1$ in this expression differ from those of $\mathcal{N}=1$ SYM.
The normalization scale $\mu_C$ in the C-scheme can be chosen arbitrarily.
Its change is performed by the finite renormalizations of the coupling constant $a_s^C \equiv \alpha_s^C/\pi$
\cite{Goriachuk:2020wyn}.
These transformations, conserving the $\beta$-function (\ref{eq:betaC}), are unambiguously described
by the only parameter $C \equiv \ln{\left(\mu_C^2/\mu^2\right)}$ (responsible for the term <<C-scheme>>),
which is similar to the analogous scale parameter, first introduced in \cite{Vladimirov:1979my}.
Scale transformations in QCD also exhibit the properties of a commutative subgroup
of renormalization group transformations \cite{Brodsky:2012ms}.

The transition to the C-scheme is performed by the help of a special finite renormalization, which is also analogous
to that converting the $\beta$-function of $\mathcal{N}=1$ SYM from the $\DRbar$ scheme one
into the exact Novikov, Shifman, Vainshtein and Zakharov $\beta$-function \cite{Jack:1996vg}.
This finite renormalization is made when the required quantum corrections are already computed at the desired order
with the particular initial renormalization prescription ($\MSbar$ in the QCD case), defined at the scale $\mu$.
The $\beta$-function of this initial scheme is defined as follows:
\begin{equation}
\beta\left(a_s\right) = \mu^2 \frac{\partial a_s\left(\mu^2\right)}{\partial\mu^2} =
- \sum_{i \geq 0} \beta_{i} {a_s}^{i+2}
\label{eq:beta-exp}
\end{equation}
(where $a_s \equiv \alpha_s/\pi$).
It determines the renormalization group evolution of the coupling constant with the change of the $\mu$ scale.
Its coefficients $\beta_i$ for $i \geq 2$ depend on the renormalization scheme choice, whereas the first two of them
are scheme-independent.
For the case of QCD evaluated in \cite{Gross:1973id,Politzer:1973fx}  $\beta_0$ coefficient and
obtained in \cite{Caswell:1974gg,Jones:1974mm,Egorian:1978zx} $\beta_1$ term read as follows:
\begin{equation}
\beta_{0} = \frac{11}{12} C_A - \frac{1}{3} T_F n_f, \quad
\beta_{1} = \frac{17}{24} {C_A}^2 - \frac{5}{12} C_A T_F n_f - \frac{1}{4} C_F T_F n_f. \label{eq:beta0beta1_QCD}
\end{equation}
Here $C_F$ and $C_A$ stand for the Casimir operator in the fundamental and adjoint representations of the
$SU\left(N_c\right)$ colour gauge group, $n_f$ is the number of quark flavours,
$T_F$ denotes the Dynkin index, which is related to the dimension $d_R$ of the
fundamental representation and the group dimension $N_A$ in the following way:
\begin{equation}
T_F N_A = C_F d_R. \label{eq:TfNa-Cfdr}
\end{equation}

Analytical representation for the coefficients of the QCD $\beta$-function in the $\MSbar$-scheme
is known up to the 5-loop approximation.
The expression for $\beta_2$, obtained in \cite{Tarasov:1980au,Larin:1993tp}, contains only rational numbers,
similarly to equations (\ref{eq:beta0beta1_QCD}).
But the $\MSbar$-scheme $\beta$-function in the subsequent orders of the PT include terms with transcendental functions,
such as the Riemann $\zeta$-function of integer argument:
\begin{equation}
\zeta_n \equiv \zeta\left(n\right) = \sum_{k=1}^{\infty} \frac{1}{k^n}.
\end{equation}
The 4-loop coefficient $\beta_3$ (see \cite{vanRitbergen:1997va,Czakon:2004bu}) is the lowest order one,
containing $\zeta_3$.
Along with this transcendency, new colour factors appear in the analytical expression for this coefficient, namely,
$d_F^{abcd}$ and $d_A^{abcd}$, representing the symmetrized traces of four generators in the fundamental and
adjoint representations respectively.
The five-loop coefficient $\beta_4$, analytically calculated in \cite{Baikov:2016tgj,Herzog:2017ohr,Luthe:2017ttc},
together with $\zeta_3$ contains $\zeta_4$ and $\zeta_5$.

The C-scheme $\beta$-function (\ref{eq:betaC}) can also be expanded into the PT series (\ref{eq:beta-exp}).
Its coefficients do not contain transcendental values.
It is not the only one remarkable feature of the C-scheme.
The terms, proportional to $\zeta_{2m}$ (for $m = 2, 3, \dots$), are canceled out in this scheme in
higher-order PT corrections to the renormalization-group invariant physical quantities, defined in Euclidean space
\cite{Jamin:2017mul}.
This cancellations of <<$\pi^2$>>-like transcendences are not occasional and related to the all-order
no-$\pi$ theorems \cite{Baikov:2018wgs}, which are also valid for generic anomalous dimension,
related to the Higgs boson quark-antiquark decay \cite{Vermaseren:1997fq}\footnote{The corresponding
generic anomalous dimension was considered previously in \cite{Gorishnii:1991zr}},
and the ones of the DIS structure functions \cite{Davies:2017hyl,Herzog:2018kwj}.
In addition, no-$\pi$ theorems permit to deduce the relations between the transcendental contributions
to the coefficients of the $\beta$-function and anomalous dimensions in arbitrary theory
with the unique coupling constant, starting from the 4-loop approximation \cite{Baikov:2018wgs,Baikov:2019zmy}.

In $SU\left(N_c\right)$ QCD the problem of fixing $\zeta_4$-contributions to the re\-nor\-ma\-li\-za\-ti\-on-group
governed quantities, defined in the Euclidean space-time region, in the fifth order of the PT in the $\MSbar$ scheme
was considered previously in \cite{Davies:2017hyl}.
We perform the similar analysis for two renormalization-group invariant physical quantities, namely Adler D-function
of electromagnetic vector currents \cite{Adler:1974gd}, related to the total cross-section of electron-positron
annihilation to hadrons and Bjorken polarized deep-inelastic lepton-nucleon scattering (DIS) sum rule
\cite{Bjorken:1969mm}.
The dependence of the Bjorken coefficient function on the squared Euclidean momentum was extracted from the experimental
data and employed for the comparison with the theoretical predictions of QCD \cite{Kotlorz:2018bxp}
(see also \cite{Deur:2018roz} for a review).
In the case of the D-function the comparison of the QCD predictions with the experimentally-motivated data was performed
in \cite{Eidelman:1998vc}.
While evaluating the QCD analytical expressions of the third order corrections to the D-function
(in the $\MSbar$ scheme) it was emphasised, that $\zeta_4$-terms appear at the intermediate steps of the computations,
but completely cancel out in the ultimate result \cite{Gorishnii:1990vf}.
Perturbative expansions for Adler D-function and Bjorken sum rule do not contain such terms
up to the fourth order of $a_s$ \cite{Baikov:2010je}.
Riemann $\zeta_4$ appears at the fifth order of the PT series for the D-function,
renormalized with the $\MSbar$ scheme.
This was shown in \cite{Jamin:2017mul} for the $SU\left(3\right)$ QCD.

In the present paper the expressions for the $\zeta_4$-contributions to Adler and Bjorken coefficient functions
in the fifth order of $a_s$ are deduced for the case of the general $SU\left(N_c\right)$ theory.
The invariance of these contributions with respect to the change of the normalization scale is shown.
Mutual cancellations of them in the original generalization of Crewther identity, first discovered at the 3-loop order
in \cite{Broadhurst:1993ru} and confirmed at the forth order of PT in \cite{Baikov:2010je}, are demonstrated\footnote{
Another form of the generalized Crewther relation was proposed in \cite{Kataev:2010du}}.
The origin of these cancellations and their relation to the ones, which occur in the conformal-invariant limit
of this identity \cite{Crewther:1972kn}, is investigated.

\section{Adler D-function and Bjorken sum rule}
Adler D-function is defined as the derivative of the polarization operator
over the squared Euclidean four-momentum $P^2$.
It is related to the total cross section $\sigma_{tot}\left(e^+e^- \to \mathrm{hadrons}\right)$:
\begin{equation}
D_A\left(a_s, \frac{P^2}{\mu^2}\right) = - 12 \pi^2 P^2 \frac{d}{d P^2} \Pi\left(a_s, \frac{P^2}{\mu^2}\right) =
P^2 \int\limits_0^\infty \frac{R\left(s\right)}{\left(P^2 + s\right)^2} \mathrm{d} s, \label{eq:RtoDint-rel}
\end{equation}
where
$R\left(s\right) \equiv \sigma_{tot}\left(e^+e^- \to \mathrm{hadrons}\right)/\sigma\left(e^+e^- \to \mu^+\mu^-\right)$.
We will investigate normalized on unity flavour non-singlet part $D^{ns}\left(a_s, P^2/\mu^2\right)$ of the D-function,
which enters the following equation:
\begin{equation}
D_A\left(a_s, \frac{P^2}{\mu^2}\right) = d_R \left(\sum_f {Q_f}^2\right) D^{ns} \left(a_s, \frac{P^2}{\mu^2}\right) +
d_R \left(\sum_f Q_f\right)^2 D^{si} \left(a_s, \frac{P^2}{\mu^2}\right),
\label{eq:Dnorm}
\end{equation}
where $d_R$ is the number of quark flavours and $Q_f$ denotes the electric charge of the quark with the flavour $f$.
The singlet corrections $D^{si}\left(a_s, P^2/\mu^2\right)$ to the D-function first appear at the third order of $a_s$
\cite{Gorishnii:1990vf,Surguladze:1990tg}.
Note, that flavour singlet terms do not contribute the full expression in the case of $n_f = 3$
due to the vanishing prefactors $\sum Q_f$.

Bjorken sum rule \cite{Bjorken:1969mm} is used for the description of deep-inelastic scattering of leptons ($l$)
on polarized nucleons.
It is proportional to the difference of the spin structure functions $g_1^{lp}$ and $g_1^{ln}$ of this process,
integrated over the spin distribution within the nucleon (proton ($p$) and neutron ($n$)):
\begin{equation}
\left|\frac{g_A}{6 g_V}\right| C_{Bjp}\left(a_s, \frac{P^2}{\mu^2}\right) =
\int\limits_0^1 \left(g_1^{lp}\left(x, P^2\right) - g_1^{ln}\left(x, P^2\right)\right)\mathrm{d} x, \label{eq:Bjp}
\end{equation}
where $g_A$ and $g_V$ denote axial and vector charges of the nucleon, correspondingly.
Bjorken coefficient function contains both the singlet and non-singlet contributions.
The former is proportional to $d_R \sum_f Q_f$.
In contrast to the D-function, the singlet part $C_{Bjp}^{si}\left(a_s, P^2/\mu^2\right)$ appears only in the
4-th order of PT \cite{Larin:2013yba,Baikov:2015tea}.
Later on we will consider non-singlet contributions to (\ref{eq:RtoDint-rel}) and (\ref{eq:Bjp}) only.

The D-function and Bjorken sum rule are renormalization-group invariants and satisfy the following relations:
\begin{eqnarray}
D^{ns}\bigg(a_s\left(\mu^2\right),~ P^2/\mu^2\bigg) &=&
 D_0^{ns}\bigg(a_{s0},~ P^2/\mu^2\bigg), \nonumber \\
C_{Bjp}\bigg(a_s\left(\mu^2\right),~ P^2/\mu^2\bigg) &=&
 C^{ns}_{Bjp0}\bigg(a_{s0},~ P^2/\mu^2\bigg).
\label{eqs:D-C_toBare}
\end{eqnarray}
The regularization scale (chosen to be equal to $\mu$) in the right-hand sides of these equations
(defined in terms of the bare coupling constant $\alpha_{s0}$) is required to restore the original momentum dimension
of the measure of loop integrals, calculated in $d = 4 - 2 \epsilon$ space-time dimensions
within the dimensional regularization (or dimensional reduction) technique.

Since $D^{ns}$ and $C_{Bjp}^{ns}$ in equation (\ref{eqs:D-C_toBare}) do not depend on the choice
of the renormalization scheme, they remain invariant to the shifts of the normalization scale $\mu$.
Therefore, their total derivatives with respect to  $\mu^2$ vanish for arbitrary $P^2$:
\begin{equation}
\mu^2\frac{d}{d \mu^2} \left(
\begin{array}{c}
D^{ns}\left(a_s, \frac{P^2}{\mu^2}\right) \\
C_{Bjp}^{ns}\left(a_s, \frac{P^2}{\mu^2}\right)
\end{array}
\right) = \left(\mu^2\frac{\partial}{\partial\mu^2} +
\beta\left(a_s\right)\frac{\partial}{\partial a_s}\right) \left(
\begin{array}{c}
D^{ns}\left(a_s, \frac{P^2}{\mu^2}\right) \\
C_{Bjp}^{ns}\left(a_s, \frac{P^2}{\mu^2}\right)
\end{array}
\right) = 0, \label{eq:D-RGeq}
\end{equation}
where the $\beta$-function is defined in (\ref{eq:beta-exp}).

$D^{ns}\left(a_s, P^2/\mu^2\right)$ and $C^{ns}_{Bjp}\left(a_s, P^2/\mu^2\right)$ are functions of the squared
Euclidean four-momentum $P^2$.
The logarithms $\ln{\left(P^2/\mu^2\right)}$, entering them, can always be restored using the renormalization group
technique.
The solutions of the renormalization group equations (\ref{eq:D-RGeq}) read:
\begin{equation}
D^{ns}\left(a_s\right) \equiv D^{ns}\left(a_s, 1\right), \quad
C^{ns}_{Bjp}\left(a_s\right) \equiv C^{ns}_{Bjp}\left(a_s, 1\right),
\end{equation}
where $a_s = a_s\left(P^2\right)$ is the QCD running coupling constant.
Then $D^{ns}\left(a_s\right)$ and $C^{ns}_{Bjp}\left(a_s\right)$ depend exclusively on $a_s\left(P^2\right)$ and
the following PT expansion is applied for them:
\begin{equation}
D^{ns}\left(a_s\right) = 1 + \sum_{r \geq 1} d_r {a_s}^r, \quad
C^{ns}_{Bjp}\left(a_s\right) = 1 + \sum_{r \geq 1} c_r {a_s}^r. \label{eqs:D-C_PTExp}
\end{equation}

In \cite{Mikhailov:2004iq} the $\beta$-expanded form for the coefficients
of these re\-nor\-ma\-li\-za\-ti\-on-group invariants was proposed.
The $\beta$-expansion for the low order coefficients of, for example, the D-function
in (\ref{eqs:D-C_PTExp}) reads:
\begin{eqnarray}
d_1 &=& d_1[0], \nonumber \\
d_2 &=& \beta_0 d_2[1] + d_2[0], \nonumber \\ 
d_3 &=& {\beta_0}^2 d_3[2] + \beta_1 d_3[0,1] + \beta_0 d_3[1] + d_3[0], \nonumber \\ 
d_4 &=& {\beta_0}^3 d_4[3] + \beta_1 \beta_0 d_4[1,1] + {\beta_0}^2 d_4[2] + \nonumber \\
&& \beta_2 d_4[0,0,1] + \beta_1  d_4[0,1] + \beta_0 d_4[1] + d_4[0]. \label{eqs:d1-d4beta-exp}
\end{eqnarray}
where the natural number at the $i$-th place in the square brackets after each coefficient
denotes the power of $\beta_{i-1}$ in the prefactor in front of it.
The similar $\beta$-expansion for the non-singlet terms of Bjorken sum rule
was first considered in \cite{Kataev:2010du}.

Flavour non-singlet Adler function and Bjorken polarized DIS sum rule in $\MSbar$ scheme satisfy
the generalized Crewther identity, which can be written down in the following form:
\begin{equation}
D^{ns}\left(a_s, P^2/\mu^2\right) C^{ns}_{Bjp}\left(a_s, P^2/\mu^2\right) = 1 + \Delta\left(a_s, P^2/\mu^2\right).
\label{eq:GenCrewterId}
\end{equation}
Its renormalization scale and scheme dependence was studied in \cite{Shen:2016dnq,Garkusha:2018mua}.
In the $\MSbar$-like schemes $\Delta\left(a_s, P^2/\mu^2\right)$ in the right hand side of (\ref{eq:GenCrewterId})
can be expressed in the following form \cite{Baikov:2010je,Broadhurst:1993ru,Gabadadze:1995ei,Crewther:1997ux}:
\begin{equation}
\Delta\left(a_s, 1\right) = \beta\left(a_s\right) K\left(a_s\right) =
\beta\left(a_s\right) \sum_{r \geq 0} K_{r} {a_s}^r. \label{eq:CBK}
\end{equation}
Hence it vanishes in the conformal invariant limit \cite{Kataev:2013vua},
reached by turning the QCD $\beta$-function into zero, in which the product of the D-function and Bjorken sum rule
equals unity.
As a consequence, scheme and scale independent coefficients $d_1$ and $c_1$ in expansions (\ref{eqs:D-C_PTExp})
have opposite signs and equal magnitudes.
Indeed, analytical expressions of these coefficients, obtained in \cite{Zee:1973sr,Appelquist:1973uz} and
\cite{Kodaira:1979ib} respectively, have the following form:
\begin{equation}
d_1 = \left(3/4\right) C_F, \qquad c_1 = - \left(3/4\right) C_F. \label{eqs:d1c1}
\end{equation}

\section{$\zeta_4$-contributions to $d_5$ and $c_5$ in $SU\left(N_c\right)$ QCD}
As it was mentioned above the 5-th order contributions to Adler D-function, proportional to $\zeta_4$
in the $\MSbar$ renormalization scheme can be restored with the help of finite renormalizations,
performing the transition to the C-scheme \cite{Jamin:2017mul}.
One can obtain similar corrections to Bjorken sum rule in the same way.
To evaluate $\zeta_4$-terms in the 5-th order contributions to $D^{ns}\left(a_s\right)$ and
$C^{ns}_{Bjp}\left(a_s\right)$ in $SU\left(N_c\right)$ QCD (i.e., $d_5^{\left(\zeta_4\right)}$ and
$c_5^{\left(\zeta_4\right)}$, see expansions (\ref{eqs:D-C_PTExp})) we use an alternative approach, based on the
effective charges method \cite{Grunberg:1982fw} (see also \cite{Kataev:1995vh} for the definite studies).
Both the finite renormalizations technique and the effective charges approach rely upon the no-$\pi$ theorem,
related to the cancellation of the terms with $\zeta_4$ in Euclidean physical quantities in the C-scheme
\cite{Baikov:2018wgs}.
This no-$\pi$ theorem guarantees that the terms with $\zeta_4$ do not appear in Adler and Bjorken functions
in the C-scheme up to the fifth order and beyond.

The transition from the particular initial renormalization scheme to another one is performed by the redefinition
of the coupling constant $a_s^\prime\left(a_s\right)$ (i.e., by the finite renormalization).
Due to renormalization group invariance (\ref{eqs:D-C_toBare}) of the considered quantities, the following equalities
are valid for them:
\begin{eqnarray}
{D^{ns}}^\prime\left({a_s}^\prime\left(a_s\right), P^2/\mu^2\right) &=& D^{ns}\left(a_s, P^2/\mu^2\right), \nonumber \\
{C^{ns}_{Bjp}}^\prime\left({a_s}^\prime\left(a_s\right), P^2/\mu^2\right) &=& C^{ns}_{Bjp}\left(a_s, P^2/\mu^2\right).
\label{eqs:D-C_trLaw}
\end{eqnarray}

Let us choose the renormalization prescription in which the perturbative expansion for Adler D-function includes
only the first-order correction:
\begin{equation}
D^{ns}_{eff}\left(a_s^{eff(D^{ns})}\right) = 1 + d_1 a_s^{eff(D^{ns})}. \label{eq:Dns-eff}
\end{equation}
The coupling constant in the the resulting effective scheme is called the $D^{ns}$-effective charge.
It is related to $a_s$ by the following finite renormalization:
\begin{equation}
a_s^{eff(D^{ns})}\left(a_s\right) = a_s + \frac{1}{d_1} \sum_{r \geq 2} d_r {a_s}^r. \label{eq:as-eff}
\end{equation}
The effective charge is defined at a different normalization scale
$\mu_{eff} = \mu \exp{\left(- d_2/\left(2 \beta_0 d_1\right)\right)}$ \cite{Krasnikov:1981rp}
and satisfies the following renormalization-group equation:
\begin{equation}
\beta^{eff(D^{ns})}\left(a_s^{eff(D^{ns})}\right) =
\mu_{eff}^2 \frac{\partial a_s^{eff(D^{ns})}\left(\mu_{eff}^2\right)}{\partial\mu_{eff}^2} =
- \sum_{i \geq 0} \beta_{i}^{eff(D^{ns})} {\left(a_s^{eff(D^{ns})}\right)}^{i+2}. \label{eq:beta-effD}
\end{equation}

The coefficients $\beta_{i}^{eff(D^{ns})}$ in this expansion for the $\beta$-function
are related to scheme invariants \cite{Stevenson:1981vj} and do not depend on the initial
gauge-invariant renormalization scheme.
For example, the expression for $\beta_4^{eff(D^{ns})}$ reads:
\begin{eqnarray}
\beta_4^{eff(D^{ns})} &=& \beta_4 - 3 \beta_3 \frac{d_2}{d_1} +
\beta_2 \left(-\frac{d_3}{d_1} + 4 \frac{{d_2}^2}{{d_1}^2}\right) +
\beta_1 \left(\frac{d_4}{d_1} - 2 \frac{d_2 d_3}{{d_1}^2}\right) + \nonumber \\
 && \beta_0 \left(3 \frac{d_5}{d_1} - 12 \frac{d_2 d_4}{{d_1}^2} - 5 \frac{{d_3}^2}{{d_1}^2} +
28 \frac{{d_2}^2 d_3}{{d_1}^3} - 14 \frac{{d_2}^4}{{d_1}^4}\right). \label{eq:beta4inv}
\end{eqnarray}
It coincides with the result of \cite{Kataev:1995vh}.

The analytical expression for $\beta_4^{eff(D^{ns})}$ does not contain terms with $\zeta_4$.
To verify this, we consider formula (\ref{eq:beta4inv}) in the C-scheme.
Neither the D-function coefficients in this formula, nor those of the $\beta$-function (\ref{eq:betaC})
do not contain $\zeta_4$ in the C-scheme.

The analytical representation for $\beta_4^{eff(D^{ns})}$ in (\ref{eq:beta4inv}) is the same,
whether it is calculated in the C-scheme or in the $\MSbar$-like one.
Let us return to the case of $\MSbar$-like renormalization scheme and find the relations,
following from equation (\ref{eq:beta4inv}).
We observe, that all the terms with $\zeta_4$, entering it, must satisfy the following equality:
\begin{equation}
0 = \beta_4^{\left(\zeta_4\right)} + \beta_0 ~ 3 \frac{d_5^{\left(\zeta_4\right)}}{d_1}. \label{eq:Beta4invDz4}
\end{equation}

This formula presumes the exact division of $d_5^{\left(\zeta_4\right)}$ by $d_1$.
Indeed, $d_1$ is proportional to $C_F$ (see equations (\ref{eqs:d1c1})), $\beta_4^{\left(\zeta_4\right)}$
doesn't contain $C_F$ in the denominator and there is no $C_F$ in $\beta_0$ (see (\ref{eq:beta0beta1_QCD})).
Moreover, due to equation (\ref{eq:Beta4invDz4}) $\beta_4^{\left(\zeta_4\right)}$ is proportional to $\beta_0$.
Hence, the $\zeta_4$-part of the 5-loop coefficient of the QCD $\beta$-function in the $\MSbar$-like schemes
is divided by $\beta_0$ without the reminder.
This statement was obtained earlier from alternative reasoning \cite{Baikov:2018wgs}, leading to the relation
between the $\zeta_4$-terms in $\beta_4$ and the $\zeta_3$-contributions to 4-loop QCD $\beta$-function:
\begin{equation}
\frac{\beta_4^{\left(\zeta_4\right)}}{\zeta_4} = - \frac{9}{8} \beta_0 \frac{\beta_3^{\left(\zeta_3\right)}}{\zeta_3}.
\end{equation}

In consistence with (\ref{eq:Beta4invDz4}), the corrections to $d_5$, proportional to $\zeta_4$,
in the $\MSbar$-like schemes obey the condition:
\begin{equation}
d_5^{\left(\zeta_4\right)} = - d_1 \frac{\beta_4^{\left(\zeta_4\right)}}{3 \beta_0}. \label{eq:d5z4}
\end{equation}
Repeating the same reasoning for the flavour non-singlet part of the Bjorken polarized DIS sum rule
one can deduce the similar equation for the $\zeta_4$-corrections to $c_5$:
\begin{equation}
c_5^{\left(\zeta_4\right)} = - c_1 \frac{\beta_4^{\left(\zeta_4\right)}}{3 \beta_0}. \label{eq:c5z4}
\end{equation}
Since $d_1 = - c_1$ (see equations (\ref{eqs:d1c1})),
the sum of $d_5^{\left(\zeta_4\right)}$ and $c_5^{\left(\zeta_4\right)}$ is equal to zero.
This property of the determined corrections is not accidental and has deep theoretical grounds, associated with
the QCD generalized Crewther identity, which will be discussed in detail further.

The coefficients of the $\beta$-function $\beta_i$ are the same for all $\MSbar$-like schemes.
For $SU\left(N_c\right)$ QCD they are known analytically up to the 5-loop approximation \cite{Herzog:2017ohr},
in contrast to yet unknown $d_5$ and $c_5$.
The part of the 5-loop $\beta$-function coefficient, proportional to $\zeta_4$,
is given by the following analytical expression:
\begin{eqnarray}
 && \beta_4^{\left(\zeta_4\right)} = \Bigg(\frac{121}{6144} {C_A}^5
- \frac{121}{128} \frac{d_A^{abcd}d_A^{abcd}}{N_A} C_A - \frac{583}{3072} {C_A}^4 T_F n_f + \nonumber \\
 && + \frac{451}{1536} C_F {C_A}^3 T_F n_f - \frac{121}{768} {C_A}^2 {C_F}^2 T_F n_f +
\frac{11}{32} \frac{d_A^{abcd}d_A^{abcd}}{N_A} T_F n_f + \nonumber \\
 && + \frac{143}{64} \frac{d_F^{abcd}d_A^{abcd}}{N_A} C_A n_f - \frac{13}{384} {C_A}^3 \left(T_F n_f\right)^2 -
\frac{59}{192} C_F {C_A}^2 \left(T_F n_f\right)^2 + \nonumber \\
 && + \frac{143}{384} {C_F}^2 C_A \left(T_F n_f\right)^2 - \frac{13}{16} \frac{d_F^{abcd}d_A^{abcd}}{N_A} T_F {n_f}^2 -
\frac{11}{16} \frac{d_F^{abcd}d_F^{abcd}}{N_A} C_A {n_f}^2 + \nonumber \\
 && \frac{7}{192} {C_A}^2 \left(T_F n_f\right)^3 + \frac{7}{96} C_F C_A \left(T_F n_f\right)^3 -
\frac{11}{96} {C_F}^2 \left(T_F n_f\right)^3 + \nonumber \\
 && +\frac{1}{4} \frac{d_F^{abcd}d_F^{abcd}}{N_A} T_F {n_f}^3 \Bigg) \zeta_4. \label{eq:beta4z4_QCD}
\end{eqnarray}

Using (\ref{eq:TfNa-Cfdr}) it is possible to deduce the following relations:
\begin{eqnarray}
\frac{d_A^{abcd}d_A^{abcd}}{N_A} C_F &=& \frac{d_A^{abcd}d_A^{abcd}}{d_R} T_F, \nonumber \\
\frac{d_F^{abcd}d_A^{abcd}}{N_A} C_F n_f &=& \frac{d_F^{abcd}d_A^{abcd}}{d_R} T_F n_f, \nonumber \\
\frac{d_F^{abcd}d_F^{abcd}}{N_A} C_F n_f^2 &=& \frac{d_F^{abcd}d_F^{abcd}}{d_R} T_F n_f^2.
\end{eqnarray}
They permit to rewrite the particular terms in (\ref{eq:beta4z4_QCD}) in a more convenient form.
It is desirable, since the gauge group structures in the PT expressions for the flavour non-singlet
D-function\footnote{
Note, that the D-function has the $d_R$ prefactor (see relation (\ref{eq:Dnorm})),
which naturally cancels with the $d_R$ in the denominator of these group structures.}
and Bjorken sum rule differ from the colour factors in the $\beta$-function.
Indeed, the convolutions $\left(d_F^{abcd}d_F^{abcd}\right)$ and $\left(d_F^{abcd}d_A^{abcd}\right)$, divided by $d_R$,
appear in the 4-th order of these expansions \cite{Baikov:2010je}.

Taking (\ref{eqs:d1c1}) into account, we deduce the following representation
for $d_5^{\left(\zeta_4\right)}$ and $c_5^{\left(\zeta_4\right)}$,
which is valid for $\MSbar$-like renormalization schemes in $SU\left(N_c\right)$ QCD:
\begin{eqnarray}
d_5^{\left(\zeta_4\right)} &=& - c_5^{\left(\zeta_4\right)} = \Bigg(-\frac{11}{2048} C_F C_A^4 +
\frac{33}{128} \frac{d_A^{abcd}d_A^{abcd}}{d_R} T_F + \frac{11}{256} C_F^3 C_A T_F n_f - \nonumber \\
 && \frac{41}{512} C_F^2 C_A^2 T_F n_f + \frac{51}{1024} C_F C_A^3 T_F n_f -
\frac{39}{64} \frac{d_F^{abcd}d_A^{abcd}}{d_R} T_F n_f - \nonumber \\
 && \frac{11}{128} C_F^3 \left(T_F n_f\right)^2 + \frac{7}{128} C_F^2 C_A \left(T_F n_f\right)^2 + \nonumber \\
 && \frac{7}{256} C_F C_A^2 \left(T_F n_f\right)^2 +
\frac{3}{16} \frac{d_F^{abcd}d_F^{abcd}}{d_R} T_F n_f^2 \Bigg) \zeta_{4}. \label{eq:d5c5zeta4}
\end{eqnarray}
Casimir operator in the adjoint and fundamental representations, Dynkin index $T_F$ and the symmetrized traces
$d_A^{abcd}$ and $d_F^{abcd}$ of the $SU\left(N_c\right)$ group can be expressed in terms of $N_c$ in the following way
\cite{vanRitbergen:1997va}:
\begin{eqnarray}
C_F = \frac{{N_c}^2 - 1}{2 N_c} = \frac{N_A}{2 N_c}, ~ C_A = d_R = N_c, ~
\frac{d_F^{abcd} d_F^{abcd}}{N_A} = \frac{{N_c}^4 - 6 {N_c}^2 + 18}{96 {N_c}^2}, && \nonumber \\
\frac{d_F^{abcd} d_A^{abcd}}{N_A} = \frac{N_c \left({N_c}^2 + 6\right)}{48}, \quad
\frac{d_A^{abcd} d_A^{abcd}}{N_A} = \frac{{N_c}^2 \left({N_c}^2 + 36\right)}{24}, \quad T_F = \frac{1}{2}. &&
\end{eqnarray}
Substituting them into (\ref{eq:d5c5zeta4}), we get:
\begin{eqnarray}
&& d_5^{\left(\zeta_4\right)} = - c_5^{\left(\zeta_4\right)} =
\zeta_{4} \Bigg(\frac{11 N_c}{4096} \Big(-72 + 71 {N_c}^2 + {N_c}^4\Big) + \frac{1}{4096 {N_c}^2} \Big(-11 + \nonumber \\
&&  148 {N_c}^2 - 132 {N_c}^4 - 5 {N_c}^6\Big) n_f +
\frac{- 61 + 77 {N_c}^2 - 37 {N_c}^4 + 21 {N_c}^6}{4096 {N_c}^3} {n_f}^2\Bigg). \label{eq:d5c5zeta4SUNc}
\end{eqnarray}
In the case of $N_c = 3$ this formula takes the following simple form:
\begin{equation}
d_5^{\left(\zeta_4\right)} = - c_5^{\left(\zeta_4\right)} = \left(\frac{2673}{512} -
\frac{1627}{4608} n_f + \frac{809}{6912} {n_f}^2\right) \zeta_{4}. \label{eq:d5c5zeta4SU3}
\end{equation}
The coincidence of this expression for $d_5^{\left(\zeta_4\right)}$ with the result,
obtained in \cite{Jamin:2017mul} for the $\zeta_4$-corrections to the D-function of $SU\left(3\right)$ QCD,
confirms the correctness of the general $SU\left(N_c\right)$ formula, given by (\ref{eq:d5c5zeta4}).

\section{$\beta$-expansion}
The application of $\beta$-expansion procedure to Adler D-function presumes that the $\beta$-function coefficients
should be distinguished in its analytical representation \cite{Mikhailov:2004iq}.
The following simple approach is suitable for making the $\beta$-expansion: the $\left(T_F n_f\right)$ combinations
ought to be absorbed by the $\beta_i$ coefficients.
In general, this procedure, applied to the 5-th order corrections to the flavour non-singlet terms, provides
the following expansion \cite{Mikhailov:2016feh}:
\begin{eqnarray} 
d_5 &=& {\beta_0}^4 d_5[4] + \beta_1 {\beta_0}^2 d_5[2,1] + {\beta_0}^3 d_5[3] + \nonumber \\
&& \beta_2 \beta_0 d_5[1,0,1] + {\beta_1}^2 d_5[0,2] + \beta_1 \beta_0 d_5[1,1] +  {\beta_0}^2 d_5[2] + \nonumber \\
&& \beta_3 d_5[0,0,0,1] + \beta_2 d_5[0,0,1] + \beta_1 d_5[0,1] + \beta_0 d_5[1] + d_5[0]. \label{eq:d5beta-exp}
\end{eqnarray}  
An analogous formula also takes place for the flavour non-singlet terms in Bjorken sum rule
$C_{Bjp}^{ns}\left(a_s, P^2/\mu^2\right)$.
Let us consider, how the $\zeta_4$-corrections (\ref{eq:d5c5zeta4}) are distributed between the terms in this
representation.

To construct the $\beta$-expansion of these corrections additional theoretical arguments are required.
We will use the naive non-Abelization (NNA) method (see, e.g., \cite{Beneke:1994qe}).
It presumes that the $\left(T_F n_f\right)^t$ structures with the highest possible degree $t$ in the analytical
expression for a particular physical quantity (or its part, distinguished by the factor $\zeta_4$, in our case)
at each order of PT should be replaced by $\left(\left(11/4\right) C_A - 3 \beta_0\right)^t$.
The maximal $n_f$ power in corrections (\ref{eq:d5c5zeta4}) equals two, then $t = 2$.
Hence, the NNA method helps to reduce considerably the number of $\beta$-expansion coefficients in (\ref{eq:d5beta-exp}),
but it leaves the first order of $\left(T_F n_f\right)$, which can be included either in $\beta_0$ or in $\beta_1$
(or in their combination).
Indeed, both these coefficients contain $\left(T_F n_f\right)$ in the power of 1 and both the terms with $\beta_0$
and $\beta_1$ may contribute the results of direct analytical calculations (made, e.g., in \cite{Melnikov:2000qh}).
The possibility of the appearance of $\beta_1$ in the $\beta$-expansion of the D-function is considered in the
Appendix.
However, it is shown, that the coefficient $d_5^{\left(\zeta_4\right)}[0,1]$ can be set to zero.
Thus, we receive the following $\beta$-expansion for $d_5^{\left(\zeta_4\right)}$ and $c_5^{\left(\zeta_4\right)}$:
\begin{eqnarray}
d_5^{\left(\zeta_4\right)} &=& d_5^{\left(\zeta_4\right)}[0] + \beta_0 d_5^{\left(\zeta_4\right)}[1] +
{\beta_0}^2 d_5^{\left(\zeta_4\right)}[2], \nonumber \\
c_5^{\left(\zeta_4\right)} &=& c_5^{\left(\zeta_4\right)}[0] + \beta_0 c_5^{\left(\zeta_4\right)}[1] +
{\beta_0}^2 c_5^{\left(\zeta_4\right)}[2]. \label{eqs:d5z4betaExp}
\end{eqnarray}
For the coefficients in these formulas the following expansion on the QCD group factors is unambigously restored:
\begin{eqnarray}
d_5^{\left(\zeta_4\right)}[0] &=& - c_5^{\left(\zeta_4\right)}[0] = \zeta_4 \Bigg(\frac{693}{2048} C_F {C_A}^4 +
\frac{99}{512} {C_F}^2 {C_A}^3 - \frac{1089}{2048} {C_F}^3 {C_A}^2 + \nonumber \\
&& \frac{33}{128} \frac{d_A^{abcd}d_A^{abcd}}{d_R} T_F - \frac{429}{256}\frac{d_F^{abcd}d_A^{abcd}}{d_R} C_A +
\frac{33}{64} \frac{d_F^{abcd}d_F^{abcd}}{d_R} n_f C_A\Bigg), \nonumber \\
d_5^{\left(\zeta_4\right)}[1] &=& - c_5^{\left(\zeta_4\right)}[1] = \zeta_4 \Bigg(-\frac{615}{1024} C_F {C_A}^3 -
\frac{339}{512} {C_F}^2 {C_A}^2 + \frac{165}{128} {C_F}^3 C_A + \nonumber \\
&& \frac{117}{64}\frac{d_F^{abcd}d_A^{abcd}}{d_R} - \frac{9}{16}\frac{d_F^{abcd}d_F^{abcd}}{d_R} n_f\Bigg), \nonumber \\
d_5^{\left(\zeta_4\right)}[2] &=& - c_5^{\left(\zeta_4\right)}[2] = \zeta_4 \left(\frac{63}{256} C_F {C_A}^2 +
\frac{63}{128} {C_F}^2 C_A - \frac{99}{128} {C_F}^3\right). \label{eqs:d5zeta4betaExpCoeff}
\end{eqnarray}
For obtaining this representation one requires first to distinguish ${\beta_0}^2$ by absorbing the terms with the highest
degree of $\left(T_F n_f\right)$ (and, thus, find $d_5^{\left(\zeta_4\right)}[2]$ and $c_5^{\left(\zeta_4\right)}[2]$).
Then it is necessary to distinguish $\beta_0$-terms ($d_5^{\left(\zeta_4\right)}[1]$ and $c_5^{\left(\zeta_4\right)}[1]$)
in the remaining expression.
All other contributions left should be included in $d_5^{\left(\zeta_4\right)}[0]$ and $c_5^{\left(\zeta_4\right)}[0]$.
Theoretical consequences of representation (\ref{eqs:d5z4betaExp}) are discussed in the next section.

\section{Conformal invariance}
For the product of the D-function and $C^{ns}_{Bjp}$ in the generalized Crewther relation (\ref{eq:GenCrewterId}),
the following statement is applicable:
\begin{itemize}
\item[ ]
{\bf Statement.} Let $A\left(a_s, P^2/\mu^2\right)$ and $B\left(a_s, P^2/\mu^2\right)$
be renormalization group invariant quantities, defined at the same scale $\mu$.
Then their product $C\left(a_s, P^2/\mu^2\right) = A\left(a_s, P^2/\mu^2\right) B\left(a_s, P^2/\mu^2\right)$
is also renormalization group invariant.
\end{itemize}
Its validity directly follows from equation (\ref{eqs:D-C_toBare}).
Note, that the inverse is not true.
Renormalization factors of two multiplicatively renormalizable quantities may differ from unity.
But they may mutually cancel in the product.
If one chooses $D^{ns}\left(a_s, P^2/\mu^2\right)$ for the role of $A$ and $C^{ns}_{Bjp}\left(a_s, P^2/\mu^2\right)$
for $B$, their product,
\begin{equation}
S\left(a_s, P^2/\mu^2\right) = 1 + \Delta\left(a_s, P^2/\mu^2\right), \label{eq:GenCrIdRHS}
\end{equation}
representing the right hand side of the generalized Crewther identity (\ref{eq:GenCrewterId}),
would be renormalization-group invariant.
Thus, it satisfies the renor\-ma\-li\-za\-ti\-on-group equation without the anomalous dimension, similar to (\ref{eq:D-RGeq}):
\begin{equation}
\mu^2\frac{d}{d \mu^2} S\left(a_s, \frac{P^2}{\mu^2}\right) = \left(\mu^2\frac{\partial}{\partial\mu^2} +
\beta\left(a_s\right)\frac{\partial}{\partial a_s}\right) S\left(a_s, \frac{P^2}{\mu^2}\right) = 0
\label{eq:Delta-RGeq}
\end{equation}

The coefficients in the perturbative expansions for the $D^{ns}$, $C_{Bjp}^{ns}$ and $S$ renormalization-group
invariants are changed with the shift of the normalization scale $\mu$.
For example, the fifth order coefficient of the D-function, in general, exhibits the following scale transformation:
\begin{eqnarray}
d_5^\prime &=& d_5 + 4 d_4 \beta_0 L + 3 d_3 \beta_1 L + 6 d_3 {\beta_0}^2 L^2 + 2 d_2 \beta_2 L + \nonumber \\
&& 7 d_2 \beta_1 \beta_0 L^2 + 4 d_2 {\beta_0}^3 L^3 + d_1 \beta_3 L + \left(3/2\right) d_1 {\beta_1}^2 L^2 +\nonumber \\
&& 3 d_1 \beta_2 \beta_0 L^2 + \left(13/3\right) d_1 \beta_1 {\beta_0}^2 L^3 + d_1 {\beta_0}^4 L^4, \label{eqs:D_mu-chng}
\end{eqnarray}
where $L = \ln{\left({\mu^\prime}^2/\mu^2\right)}$.
The $\zeta_4$-terms (\ref{eq:d5c5zeta4}), according to this equation, receive no corrections with the change of $\mu$,
since only $d_5$ in the right hand side of this expression contains $\zeta_4$.
Indeed, consider $d_5^{\left(\zeta_4\right)}$ and $c_5^{\left(\zeta_4\right)}$, given by (\ref{eq:d5z4}) and
(\ref{eq:c5z4}) respectively.
The $\beta$-function coefficients $\beta_4^{\left(\zeta_4\right)}$ and $\beta_0$ in these equations are not affected
by the change of the normalization scale.
The first order coefficients of the D-function and Bjrken sum rule, presented in (\ref{eqs:d1c1}),
are also scale-independent.
Therefore, the terms proportional to $\zeta_4$ in $d_5$ and $c_5$ are not changed with the shift of $\mu$ and
the expansion (\ref{eq:d5c5zeta4}) stays valid in the class of $\MSbar$-like renormalization schemes.

Similarly to (\ref{eqs:D-C_PTExp}), one can write the solution of the renormalization-group equation
(\ref{eq:Delta-RGeq}) in the form of the perturbative series:
\begin{equation}
S\left(a_s, 1\right) = 1 + \beta\left(a_s\right) K\left(a_s\right) = 1 + \sum_{r \geq 1} S_r {a_s}^r,
\label{eqs:Delta_PTExp}
\end{equation}
where (\ref{eq:CBK}) was taken into account.
Due to the generalized Crewther identity (\ref{eq:GenCrewterId}), the coefficients in this equation
are related to those in (\ref{eqs:D-C_PTExp}) by the following formula (for $r \geq 1$):
\begin{equation}
S_r = d_r + c_r + \sum_{k=1}^{r-1} d_k c_{r-k}.
\end{equation}
For $r \leq 5$ these coefficients do not contain $\zeta_4$-terms.
Indeed, in accordance with (\ref{eq:d5c5zeta4}), $d_5^{\left(\zeta_4\right)}$ and $c_5^{\left(\zeta_4\right)}$
mutually cancel in the right hand side of this formula, and the previous order coefficients don`t include $\zeta_4$.
Thus, the $\zeta_4$-corrections do not appear in the right hand side of the generalized Crewther identity
(\ref{eq:GenCrewterId}) in the $\MSbar$-like renormalization schemes up to the fifth order of PT.

This cancellation of $\zeta_4$-terms in (\ref{eq:GenCrewterId}) is related to the conformal invariant limit,
in which this identity was originally deduced \cite{Crewther:1972kn}.
In this limit only unity remains in its right hand side (\ref{eq:GenCrIdRHS}) and all the coefficients $S_r$
(for $r \geq 1$) are equal to zero.
This can be formally reached by setting the $\beta$-function coefficients to zero, for example:
\begin{equation}
\left. S_5\right\arrowvert_{\beta_i \to 0} =
d_5[0] + d_4[0] c_1[0] + d_3[0] c_2[0] + d_2[0] c_3[0] + d_1[0] c_4[0] + c_5[0] = 0,
\end{equation}
where the $\beta$-expanded form of the coefficients $d_r$ and $c_r$ is employed
(see (\ref{eqs:d1-d4beta-exp}) and (\ref{eq:d5beta-exp})).
The $\zeta_4$-corrections to $S_5$ are contained only in $d_5$ and $c_5$ and mutually cancel in this sum,
since $d_5^{\left(\zeta_4\right)}[0] = - c_5^{\left(\zeta_4\right)}[0]$
(see expressions (\ref{eqs:d5zeta4betaExpCoeff})).
Thus, the cancellation of these terms is consistent with the conformal limit of the generalized Crewther identity.

The 15-parametric group of conformal transformations includes space-time translations, Lorenz boosts and rotations,
reflections, dilatations and special conformal transformations.
Particular equations may respect only some of these symmetries.
Thus, the $\zeta_4$-terms in non-singlet Adler and Bjorken functions are invariant to scale transformations (dilatations
in the four-momentum space).
The parts of these terms, represented by $d_5^{\left(\zeta_4\right)}[0]$ and $c_5^{\left(\zeta_4\right)}[0]$ in
$\beta$-expansions (\ref{eqs:d5z4betaExp}), respect the conformal limit of the generalized Crewther identity
(\ref{eq:GenCrewterId}), and, for this reason, mutually cancel when substituted into $S_5$.
The remaining terms, corresponding to $d_5^{\left(\zeta_4\right)}[1]$, $d_5^{\left(\zeta_4\right)}[2]$,
$c_5^{\left(\zeta_4\right)}[1]$ and $c_5^{\left(\zeta_4\right)}[2]$,
cancel in this identity only due to their scale invariance.
Thus, mutual cancellation of the particular contributions to Adler D-function and Bjorken coefficient function,
observed after their substitution into (\ref{eq:GenCrewterId}), indicates the scale invariance of these contributions.

\section{Conclusions}
It has been shown in this paper that in the 5-th order of $a_s$ the contributions with $\zeta_4$ to Adler D-function
and Bjorken polarized deep-inelastic scattering sum rule can be fixed by the transition to the C-scheme.
The expression for these contributions in the $\MSbar$-like renormalization schemes has been presented in the form
of the expansion in QCD colour factors (see formula (\ref{eq:d5c5zeta4})).
It generalizes the analogous expression, obtained in the case of $SU(3)$ QCD in \cite{Jamin:2017mul}.

It has been demonstrated, that the three-parametric $\beta$-expansions (\ref{eqs:d5z4betaExp})
for the obtained $\zeta_4$-corrections to $D^{ns}$ and $C_{Bjp}^{ns}$ can be unambiguously constructed.
The analytical expressions for the coefficients of the corresponding $\beta$-expansions are given by
(\ref{eqs:d5zeta4betaExpCoeff}).
It has been shown that these corrections mutually cancel in the 5-th order of PT, when substituted into the generalized
Crewther identity (\ref{eq:GenCrewterId}), in consistence with their scale invariance.
Moreover, the cancellation of the free $\beta$-expansion coefficients $d_5^{\left(\zeta_4\right)}[0]$ and
$c_5^{\left(\zeta_4\right)}[0]$ agrees with the conformal invariant limit of this identity.

\section*{Acknowledgments}
The authors are grateful to V.S.~Molokoedov for discussions and to K.V.~Ste\-pa\-nyantz for reading the manuscript and
useful comments.

\section*{Appendix}
The naive non-Abelization approach assumes that $\left(T_F n_f\right)^t$-terms in the analytical expressions
of the flavour non-singlet Adler D-function and Bjorken DIS sum rule (or their parts,
distinguished from other corrections for some reason, e.g. due to the factor $\zeta_4$) with the highest possible $t$
should be absorbed by $\left(\beta_0\right)^t$.
Thus the terms, proportional to ${\beta_0}^2$ in (\ref{eqs:d5z4betaExp}),
are restored unambiguously by using the NNA approach.
For the same reason the terms, containing the contractions $d_F^{abcd}d_F^{abcd}$, $d_F^{abcd}d_A^{abcd}$ and
$d_A^{abcd}d_A^{abcd}$, can be unambiguously distributed between the coefficients $d_5^{\left(\zeta_4\right)}[0]$ and
$d_5^{\left(\zeta_4\right)}[1]$ in (\ref{eq:d5beta-exp}) within the NNA procedure,
since the highest degree of $\left(T_F n_f\right)$ in these terms equals unity.

The NNA approach permits to write the following $\beta$-expansion for the terms of the D-function,
containing only $C_F$, $C_A$ and $\left(T_F n_f\right)$ (we will mark them by a tilde ($~\tilde{ }~$):
\begin{eqnarray}
\tilde{d}_5^{\left(\zeta_4\right)} &=& \tilde{d}_5^{\left(\zeta_4\right)}[0] +
\beta_0 \tilde{d}_5^{\left(\zeta_4\right)}[1] + \beta_1 \tilde{d}_5^{\left(\zeta_4\right)}[0,1] +
{\beta_0}^2 \tilde{d}_5^{\left(\zeta_4\right)}[2]. \label{eqs:d5z4betaExp1}
\end{eqnarray}
The $\beta$-expansion coefficients in this expression can be represented as the following sums:
\begin{eqnarray}
\tilde{d}_5^{\left(\zeta_4\right)}[0] &=& \sum_{a=0}^{4} X_7[5-a, a] {C_F}^{5-a} {C_A}^a, \nonumber \\
\tilde{d}_5^{\left(\zeta_4\right)}[1] &=& \sum_{a=0}^{3} X_6[4-a, a] {C_F}^{4-a} {C_A}^a, \nonumber \\
\tilde{d}_5^{\left(\zeta_4\right)}[2] &=& \sum_{a=0}^{2} X_3[3-a, a] {C_F}^{3-a} {C_A}^a, \nonumber \\
\tilde{d}_5^{\left(\zeta_4\right)}[0,1] &=& \sum_{a=0}^{2} X_5[3-a, a] {C_F}^{3-a} {C_A}^a. \label{eq:d5c5z4_Xn}
\end{eqnarray}
This form of the coefficients is obtained from the analysis of the $SU\left(N_c\right)$ group structures,
entering the right hand sides of equations (\ref{eqs:d5z4betaExp1}), and consistent with another common approaches
to the construction of the $\beta$-expansion.
The similar analysis reveals twelve different structures in $\tilde{d}_5^{\left(\zeta_4\right)}$,
given by the following formula:
\begin{equation}
\tilde{d}_5^{\left(\zeta_4\right)} =
\sum_{t=0}^{2}\sum_{a=0}^{4-t} F_5[5-a-t, a, t] {C_F}^{5-a-t} {C_A}^a {\left(T_F n_f\right)}^t. \label{eq:d5z4_CfCaTfnf}
\end{equation}
Note, that only seven of these structures are represented in (\ref{eq:d5c5zeta4}), all the remaining ones equal zero:
\begin{equation}
F_5[5,0,0] = F_5[4,1,0] = F_5[3,2,0] = F_5[2,3,0] = F_5[4,0,1] = 0.
\end{equation}

All the $F_5[f,a,t]$ and $X_n[f,a]$ parameters are rational numbers, multiplied by $\zeta_4$.
The values of $F_5[f,a,t]$ can be restored from (\ref{eq:d5c5zeta4}), in contrast to fifteen unknown parameters
in (\ref{eq:d5c5z4_Xn}), represented by $X_7[f,a]$, $X_6[f,a]$, $X_3[f,a]$ and $X_5[f,a]$.
Substituting (\ref{eq:d5z4_CfCaTfnf}) and (\ref{eq:d5c5z4_Xn}) into (\ref{eqs:d5z4betaExp1}) and equating
the coefficients with different colour structures leads to the system of twelve linear algebraic equations\footnote{The
analogous approach to the construction of the $\beta$-expansion, based on the system of linear equations,
was used in \cite{Kataev:2010du,Cvetic:2016rot}.}.
For explicitly writing it the following representation of the $\beta$-function coefficients is required:
\begin{eqnarray}
\beta_0 &=& \beta_0[0,1,0] C_A + \beta_0[0,0,1] T_F n_f, \nonumber \\
\beta_1 &=& \beta_1[0,2,0] {C_A}^2 + \beta_1[0,1,0] C_A T_F n_f + \beta_1[1,0,1] C_F T_F n_f. \label{eq:Beta_toColorF}
\end{eqnarray}
The resulting system is underdetermined since it consists of 12 equations and includes 15 variables.
Hence, the dimension of the solution space of the corresponding homogeneous system is not less than $15 - 12 = 3$
(and equals to $3$, which can be explicitly verified).
For the role of three free parameters in it one can choose $X_5[f,a]$, determining the
$\tilde{d}_5^{\left(\zeta_4\right)}[0,1]$-coefficient.

The remaining 12 variables (namely, $X_7[f,a]$, $X_6[f,a]$ and $X_3[f,a]$) satisfy the system of 12 linear equations.
For arbitrary values of the free parameters this system is unambiguously solved. 
Thus, with the given $\tilde{d}_5^{\left(\zeta_4\right)}[0,1]$, all other $\beta$-expansion coefficients
in (\ref{eqs:d5z4betaExp1}) can be found.
In consistence with the NNA procedure, the $X_3[f,a]$ variables (related to $\tilde{d}_5^{\left(\zeta_4\right)}[2]$)
are fixed by this system and do not include $X_5[f,a]$ free parameters:
\begin{equation}
X_3[1,2] = \frac{F_5[1,2,2]}{\left(\beta_0[0,0,1]\right)^2}, \quad
X_3[2,1] = \frac{F_5[2,1,2]}{\left(\beta_0[0,0,1]\right)^2}, \quad
X_3[3,0] = \frac{F_5[3,0,2]}{\left(\beta_0[0,0,1]\right)^2}.
\end{equation}
Except for $X_7[5,0]$, which turns out to be zero, other variables depend on $X_5[f,a]$, for instance:
\begin{eqnarray}
&& X_6[3,1] = \frac{F_5[3,1,1]}{\beta_0[0,0,1]} - 2 \frac{\beta_0[0,1,0]}{\left(\beta_0[0,0,1]\right)^2} F_5[3,0, 2] -
\frac{\beta_1[0,1,1]}{\beta_0[0,0,1]} X_5[3,0] - \nonumber \\
&& \frac{\beta_1[1,0,1]}{\beta_0[0,0,1]} X_5[2,1], \qquad
X_6[4,0] = - \frac{\beta_1[1,0,1]}{\beta_0[0,0,1]} X_5[3,0].
\end{eqnarray}
Obviously, it is convenient to choose the free parameters to be zero,
because in this case the $\tilde{d}_5^{\left(\zeta_4\right)}[0,1]$ in (\ref{eqs:d5z4betaExp1}) completely vanishes.
This choice leads to representation (\ref{eqs:d5z4betaExp}).

Moreover, this choice can be theoretically justified by the generalization of the method,
used in \cite{Cvetic:2016rot}, to the fifth order of PT.
To avoid the description of the complicated methods we give a simple explanation for setting
$\tilde{d}_5^{\left(\zeta_4\right)}[0,1]$ to zero.
It should be noticed that corrections (\ref{eq:d5c5zeta4}) contain the $\zeta_4$-factor.
Assuming the presence of the $\beta_i$ coefficients with $i \geq 1$ in the $\beta$-expansion for these corrections,
one should expect the existence of similar terms with $\zeta_4$, multiplied by $\beta_0$, in the previous PT orders.
Since there is no $\zeta_4$ in the expression for the D-function up to the 4-th order of PT
(it appears in $O\left({a_s}^5\right)$ for the first time), it is reasonable to use only the lowest order coefficient
of the $\beta$-function in the considered $\beta$-expansion.

\begin{flushleft}

\end{flushleft}
\end{document}